\documentclass[a4paper,11pt]{article}
\usepackage{PoS}
\usepackage{lineno}
\usepackage{amsmath}
\usepackage{mathtools}
\usepackage{amssymb}
\usepackage{graphicx}
\usepackage{rotating}
\usepackage{float}
\usepackage{wrapfig}
\usepackage{enumerate}
\usepackage{subcaption}
\usepackage{graphics}
\usepackage{enumitem}
\title{A Time-Variability Test for Candidate Neutrino Sources Observed with IceCube}
 \ShortTitle{Time-Variability Test for Candidate Neutrino Sources}

\author{The IceCube Collaboration \\{\normalsize \normalfont(a complete list of authors can be found at the end of the proceedings)}}

\emailAdd{pdave@gatech.edu}

\abstract{Recent studies with IceCube have shown signs of a time-integrated flux of astrophysical neutrinos from point-like sources such as TXS 0506+056 and NGC 1068. Time-variability of this neutrino emission from TXS 0506+056 has been studied extensively by assuming a temporal profile of the possible flare(s) or searching for temporal neutrino correlation with other electromagnetic counterparts. However, experimental evidence of the temporal profile of an astrophysical neutrino signal, besides the TXS 0506+056 source, remains lacking. In this study, we present a new KS-test based method for investigating time-variability. This new method complements the existing time-dependent search methods with a test for arbitrary time-variability, independent of an assumed temporal profile or electromagnetic counterpart. Additionally, this method provides a diagnostic tool for characterizing point-like source candidates in IceCube by distinguishing variable from steady neutrino emission and we show results of applying this method to a small catalog of candidate blazars.

\vspace{4mm}
{\bfseries Corresponding authors:}
Pranav Dave$^{1*}$\\
{$^{1}$ \itshape Georgia Institute of Technology}\\
$^*$ Presenter

\FullConference{37$^{\rm{th}}$ International Cosmic Ray Conference (ICRC 2021)\\
		July 12th -- 23rd, 2021\\
		Online -- Berlin, Germany}

}

\begin{document}
\maketitle

\section{Introduction}

IceCube is a cubic-kilometer neutrino detector deployed deep in the ice at the geographic South Pole \cite{Aartsen:2016nxy}. Reconstruction of the direction, energy and flavor of the neutrinos relies on the optical detection of Cherenkov radiation produced in the interactions of neutrinos in the surrounding ice or the nearby bedrock. Recent studies \cite{TXS2018,Tessa2020} have found hints of neutrino emission from blazar TXS 0506+056 as well as the Seyfert galaxy NGC 1068. In this era of multimessenger astronomy, IceCube can provide valuable insight into the sources of cosmic rays and high-energy gamma rays as demonstrated by the follow-up of alert IC170922A \cite{TXS2018_mm} and its association with the blazar TXS 0506+056. 

Detection of astrophysical neutrinos from known gamma-ray emitters can be the "smoking gun" evidence for a hadronic component in their $\gamma$-ray spectrum. The two objects of interest, a blazar (TXS 0506+056) and a nearby Seyfert galaxy (NGC 1068), show differing features in their $\gamma$-ray spectrum. NGC 1068 has shown no signs of $\gamma$-ray variability \cite{FermiGeV_2012} while $\gtrsim$ TeV neutrinos can be expected from an AGN outflow model \cite{AGN_outflow_2016}. Additionally, more than 50\% of the bright AGN Fermi LAT sample (LBAS) was found to be variable with a high significance \cite{FermiVariability_2010}. More interestingly, $\gamma$-ray variability in LBAS blazars can be perturbations in mostly steady emission or a series of possibly overlapping flares, which can be understood by random walk processes in turbulent flow in the jet or mass accretion avalanches. Neutrino variability, on the other hand, and its correlation with $\gamma$-ray variability is not well understood. For instance, TXS 0506+056 shows low $\gamma$-ray variability around the archival IceCube neutrino flare recorded in 2014-2015 \cite{Garrappa_2019}. 

Previous IceCube analyses have employed a time-dependent search with a model dependent flare hypothesis, as well as lightcurve correlation searches with observations from Fermi LAT \cite{i3timedep:2015} \cite{TXS2018}. In this work, we present a new method to test model independent time variability for astrophysical signal and background. This analysis is a single (steady) hypothesis test, so there is no null hypothesis. Here, we present the sensitivity of this test to a class of single and double flares. Additionally, we apply this test to 4 most-significant sources from the 10-year time-integrated neutrino point source search \cite{Tessa2020} as an \emph{a posteriori} check using 7.5 years of IceCube data optimized for time-dependent studies.

\section{Method}
\begin{figure}[!htb]
    \center
    \includegraphics[width=0.7\textwidth]{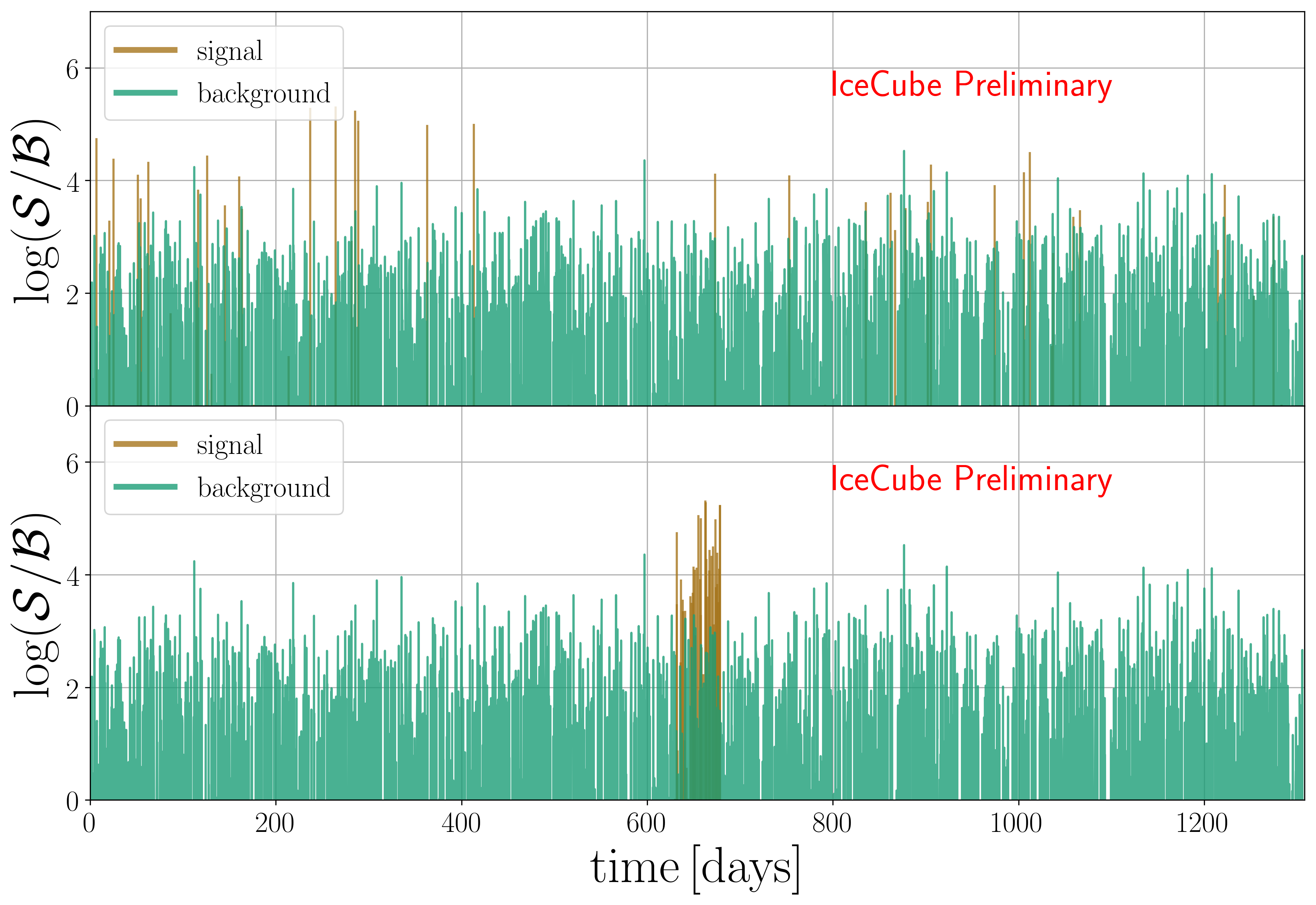}
    \caption{The log of ratio of signal and background probabilities plotted for a selection of IceCube events randomized in right ascension. Both (top and bottom) have signal events injected from Monte Carlo, however bottom plot shows a box signal temporal profile in contrast to a steady temporal profile in the upper plot.}
    \label{fig:1}
\end{figure}

In these proceedings, we use 7.5 years of track-like events \cite{realtime:2016} used by IceCube's alert system. This dataset is optimized to reduce the background of atmospheric neutrinos and atmospheric muons and consists of mostly track-like events with a median angular resolution $< 1^\circ$. Events are selected in a 10$^\circ \times 10^\circ$ box around the candidate source location. A time-integrated fit using an unbinned maximum-likelihood ratio method is performed for these events\cite{Tessa2020}. This results in two best-fit parameters describing the time-integrated astrophysical neutrino flux at this location: $n_\mathrm{fit}$ and $\gamma_\mathrm{fit}$, where $n$ is the number of signal (astrophysical) neutrinos and $\gamma$ is the spectral index of an unbroken power-law neutrino flux $\propto E^{-\gamma}$. Each event is weighted with a ratio of signal over background probability using their reconstructed arrival direction and (reconstructed) energy. The signal ($\mathbf{S}_i$) and background ($\mathbf{B}_i$) PDF can be written as the product of the spatial and energy PDFs respectively, such that the subscript $i$ refers to each event in the data sample. The background PDF is constructed from the declination-dependent experimental data, while assuming a uniform distribution in right ascension. The spatial component of the signal PDF uses a gaussian point-spread function while the energy component is the aforementioned unbroken power-law with declination dependence. The ratio of these probabilities, $\mathbf{S}/\mathbf{B}$, can be used as a diagnostic for each event. A time-series of $\mathbf{S}/\mathbf{B}$ per event is shown in Fig. \ref{fig:1} for data randomized in right ascension (background) as well as Monte-Carlo-based signal events with uniform and clustered arrival times. High-energy events close to the source have higher $\mathbf{S}/\mathbf{B}$, however the discriminating power of this ratio diminishes for a time-integrated fit with a steeper spectrum.

\begin{figure}[!htb]
    \center
    \includegraphics[width=0.7\textwidth]{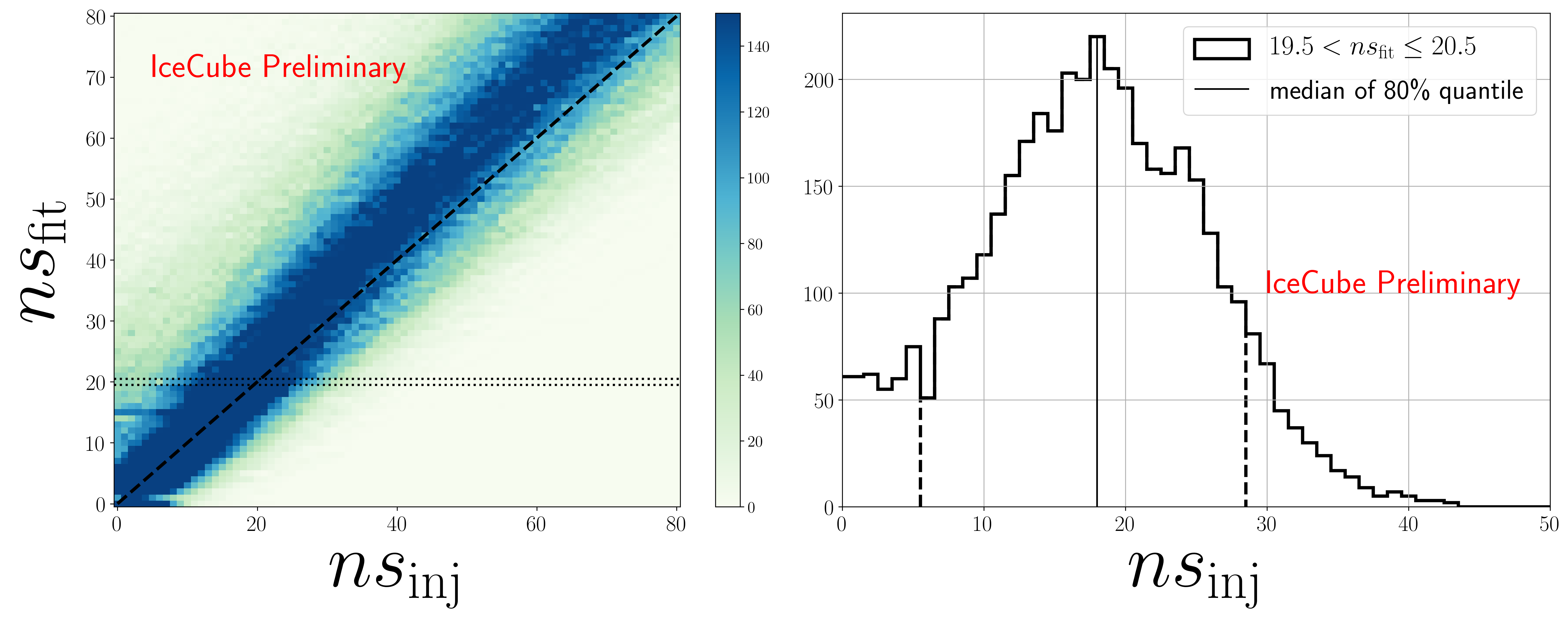}
    \caption{A distribution of injected v/s fitted number of signal events for $\gamma_\mathrm{inj}=2$ at declination $\delta=5.69^\circ$. The dashed line represents the case where all injections are fitted exactly. The two dotted horizontal lines are used to take a slice which is shown on the right, which is a distribution of all injections which fit the number of signal events near $n_\mathrm{fit}$ . The median of the middle 80\% quantile here is used to calculate $N_\mathrm{ev}$.}
    \label{fig:2}
\end{figure}

\begin{figure}[!htb]
    \center
    \includegraphics[width=0.7\textwidth]{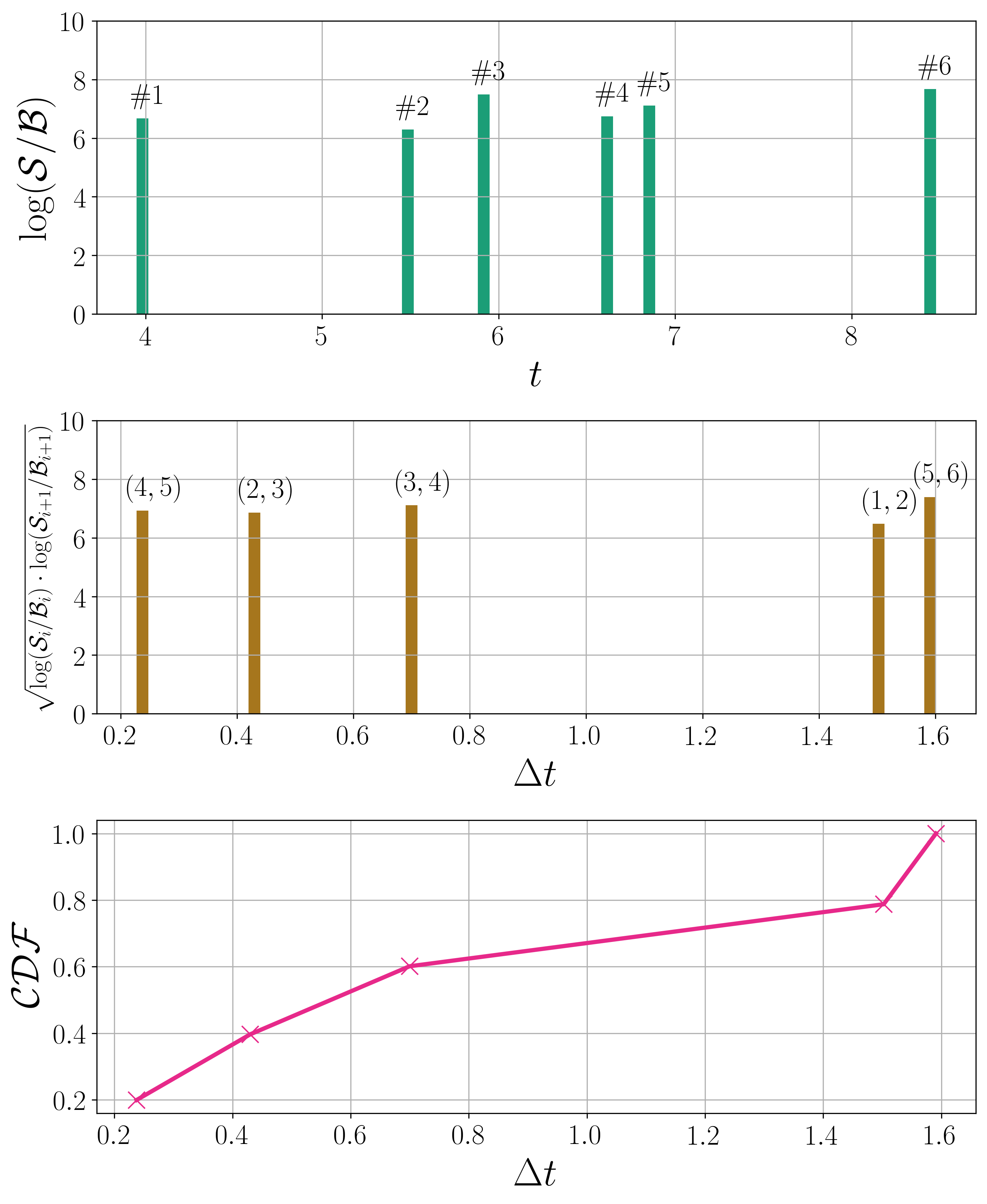}
    \caption{Example construction of the weighted CDF for the highest $\mathbf{S}/\mathbf{B}$ event pairs}
    \label{fig:3}
\end{figure}

After performing the time-integrated fit around the candidate source location, the highest $\mathbf{S}/\mathbf{B}$ events are selected based on a pre-computed injected fit-bias distribution as shown in Fig. \ref{fig:2}. This additional step is performed to minimize the bias in fitting the number of signal events. This  A weighted CDF for each consecutive event pair is constructed from these selected events based on the time difference between consecutive events and the geometric mean of the logarithm of the $\mathbf{S}/\mathbf{B}$ per-event probability ratio as the weights as shown in Fig. \ref{fig:3}. That is, each event pair is prescribed a weight, $w_i=\sqrt{\log{\left(\mathbf{S}_i/\mathbf{B}_i\right)} \times \log{\left(\mathbf{S}_{i+1}/\mathbf{B}_{i+1}\right)}}$, and a time-difference of $\Delta t = t_{i+1} - t_i$ based on their arrival times. These event pairs are then normalized, $\sum w_i = 1$, and compared against the empirical distribution function for a steady signal and background.

\begin{figure}[!htb]
    \center
    \includegraphics[width=0.7\textwidth]{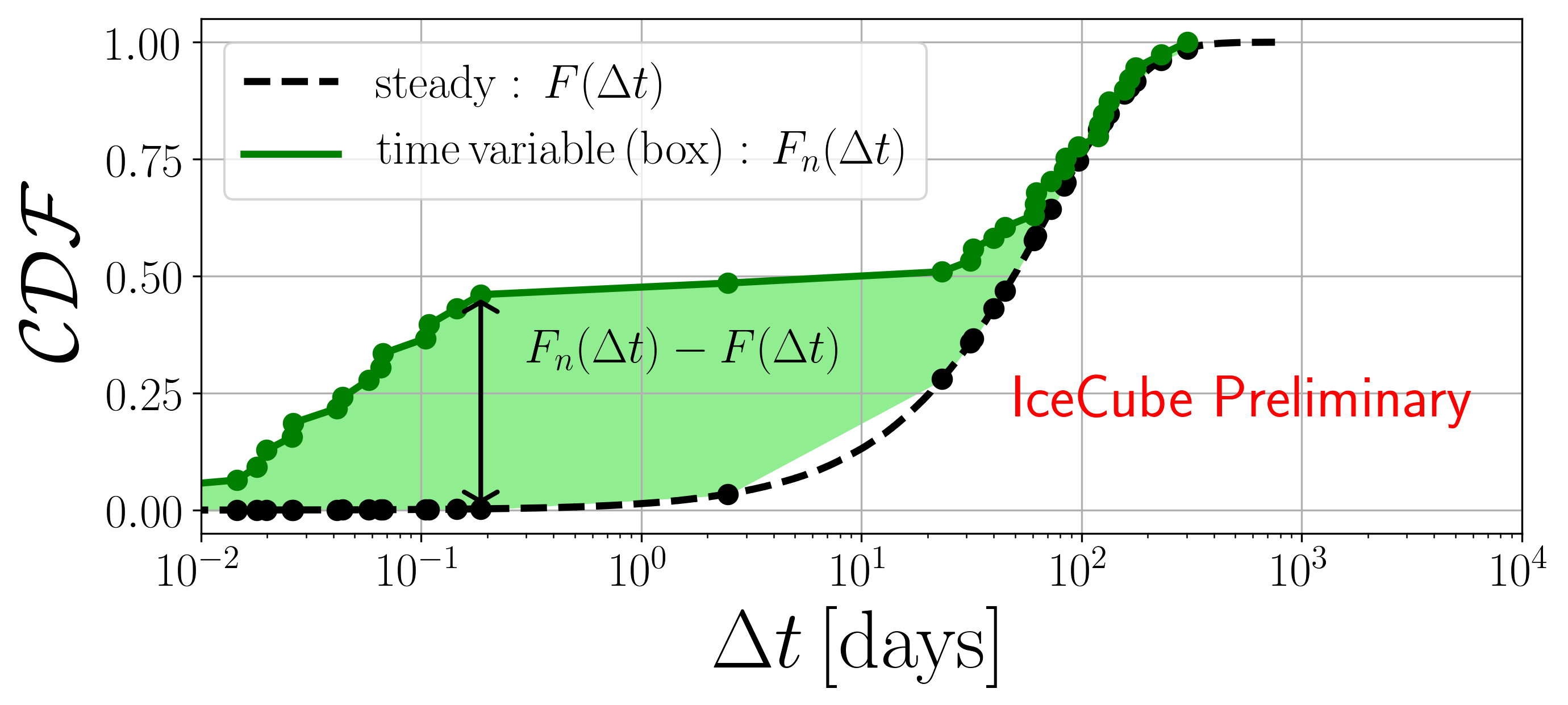}
    \caption{The weighted CDFs of consecutive pairs used to calculate the test statistic.}
    \label{fig:4}
\end{figure}

The weighted CDF can be compared to an equal-weighted hypothesis which would correspond to steady signal and background. We use a Cramer-von Mises test to calculate the test-statistic, $TS$, by integrating the distance between the weighted CDF calculated from data ($F_n(\Delta t)$) and the empirical distribution function ($F(\Delta t)$) which represents the steady hypothesis being tested as shown in Fig. \ref{fig:4}. Here, the time difference between consecutive events for the highest $\mathbf{S}/\mathbf{B}$ events is $\Delta t$ and the test statistic $TS$ used for hypothesis testing can be calculated using: 

\begin{align}
    TS &= \sqrt{N_\mathrm{ev}\,\int^1_0\,\left(F(\Delta t)-F_n(\Delta t)\right)^2\,dF}
\end{align}

\begin{figure}[!htb]
    \center
    \includegraphics[width=0.7\textwidth]{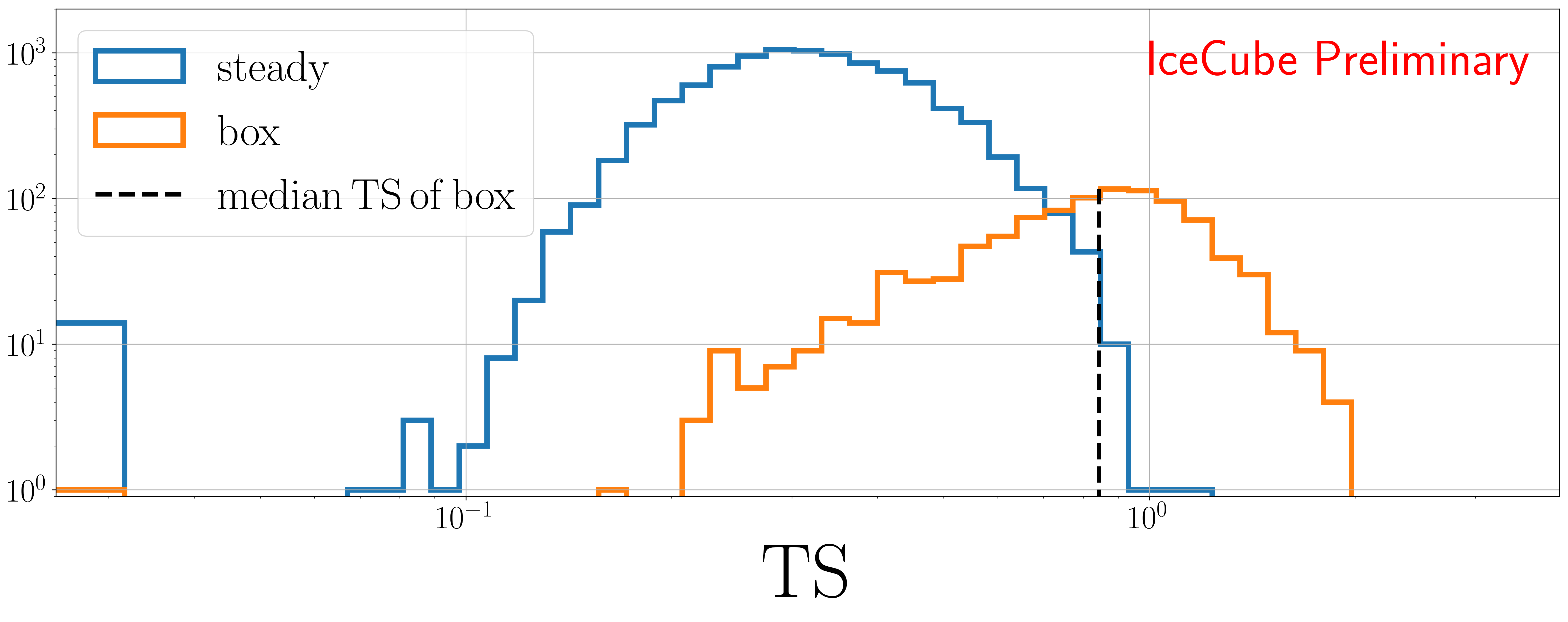}
    \caption{Test statistic distributions for time variability using injected steady and box signals. Both distributions have 20 signal event injections at declination $\delta=23.5^\circ$ and injected spectral index of $\gamma_\mathrm{inj}=2$, while the box length is 300 days. The median of the box TS is used to calculate a p-value under the steady TS, which in this case yields $2.97\sigma$ significance.}
    \label{fig:5}
\end{figure}

A set of pseudo-experiments generated from injecting signal events from Monte-Carlo in right ascension scrambled data provides a $TS$ distribution. In order to inspect the statistical variation of this $TS$ at a given declination, we fix the poisson mean of the injected signal strength ($n_\mathrm{inj}$), spectrum ($\gamma_\mathrm{inj}$), as well as the temporal profile. We compare the $TS$ distributions for a time-variable injected signal against the $TS$ distribution for a steady injection. The TS distribution associated with the time-variable signal has a median significance, calculated by comparing to the TS distribution of the steady case as shown in Fig. \ref{fig:5}. 

\begin{figure}[!htb]
    \center
    \includegraphics[width=0.7\textwidth]{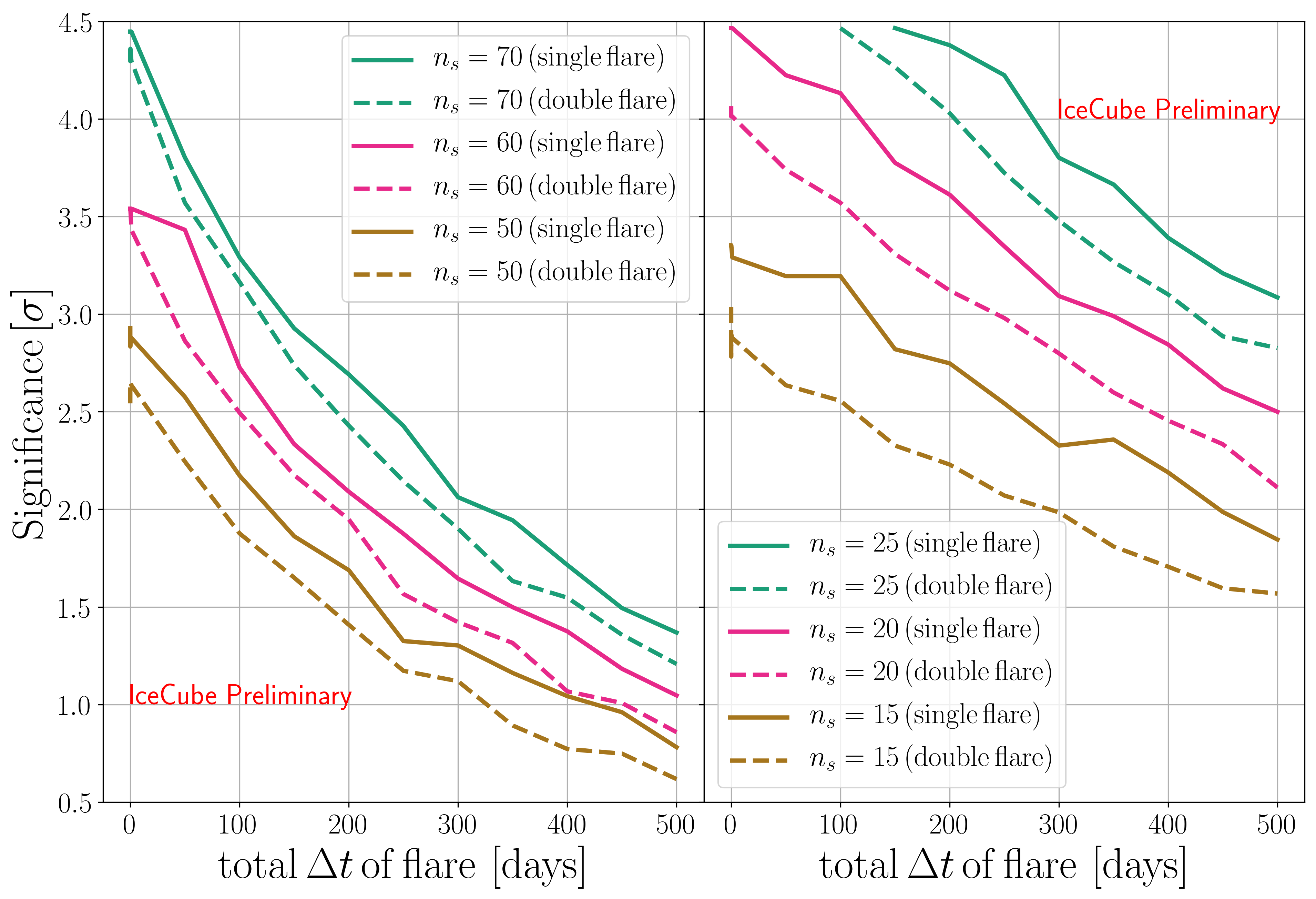}
    \caption{Single (solid) and double (dashed) flare significance for time variability in IceCube using injected signal. Double flares retain the cumulative signal in order to compare to a single flare, so that each double flare is a single flare split equally. Left:  Soft signal flare injection at declination $\delta=-0.01^\circ$, $\gamma_\mathrm{inj}=3.25$. Right: Hard signal flare injection at $\delta=5.69^\circ$, $\gamma_\mathrm{inj}=2$.}
    \label{fig:6}
\end{figure}

Sensitivities are calculated using the aforementioned significance to box-shaped flares of varying widths. As the width of the injected box-flare increases, the temporal profile naturally asymptotes to the steady hypothesis. For stronger signal strength (higher $n_\mathrm{inj}$) or harder spectra (lower $\gamma_\mathrm{inj}$), the significance of the injected box-flare increases because the discriminating power of the $\mathbf{S}/\mathbf{B}$ increases and the highest $\mathbf{S}/\mathbf{B}$ events are less likely to be contaminated by the background of atmospheric neutrinos. The sensitivity of this method to hard and soft spectra are shown in Fig. \ref{fig:6}.

\newpage

\section{Results}

A source list was constructed based on the four most significant sources from the 10 year time-integrated catalog search \cite{Tessa2020}. The data sample used here is 7.5 years and applies a slightly different event selection criteria \cite{realtime:2016}. All 4 sources are found to be consistent with steady emission as shown below:

\begin{table}[H]
    \centering
    \begin{tabular}[c]{| c | c | c | c |}
    \hline
    Name & $\alpha\,[\mathrm{deg}]$ & $\delta\,[\mathrm{deg}]$ & $p$\\
    \hline
    TXS 0506+056 & 77.35 & 5.7 & 0.62 \\
    NGC 1068 & 40.67 & -0.01 & 0.4 \\
    PKS 1424+240 & 216.76 & 23.8 & 0.53 \\
    GB6 J1542+6129 & 235.75 & 61.5 & 0.34 \\
    \hline
    \end{tabular}
    \caption{Pre-defined source-list to search for time variable neutrino emission. Source coordinates (J200) from 4FGL are provided with the pre-trial p-value.}
    \label{tab:1}
\end{table}

At first glance, the TXS 0506+056 time-variability result may seem to contradict the archival neutrino flare found in \cite{TXS2018}. However, Fig. \ref{fig:6} tells us that a signal excess of 15 events over a flaring period of ~150 days is below the $3 \sigma$ discovery threshold. Secondly, we know signal events like IC170922A which triggered the multi-messenger follow-up \cite{TXS2018_mm}, would significantly alter the model box-flare shape tested here and would likely reduce the power of this test due to more steady signal events in addition to the flare. Finally, based on Monte-Carlo signal injections, we estimate a 20\%-50\% contamination in the final event selection used in this method due to background (atmospheric) events at the declination of $TXS 0506+056$ for a simulated neutrino signal similar to the one reported \cite{TXS2018}. 

\section{Conclusions}

We introduce a method to detect time-variable astrophysical neutrino signal in IceCube. The method relies on inspecting the consecutive event-pair distribution around a candidate source and tests the steady hypothesis. This method does not assume a temporal shape for an astrophysical neutrino signal seen in IceCube and therefore has no null hypothesis, and is a single hypothesis test. We present the sensitivity of this method to box-shaped single or double flares. Using 7.5 years of track-like events used by the IceCube realtime alert system \cite{realtime:2016} we test 4 sources for time-variability, namely: NGC 1068, TXS 0506+056, PKS 1424+240, and GB6 J1542+6129. We find all 4 sources to be consistent with the steady hypothesis.
\bibliographystyle{ICRC}
\bibliography{references}
\clearpage
\section*{Full Author List: IceCube Collaboration}




\scriptsize
\noindent
R. Abbasi$^{17}$,
M. Ackermann$^{59}$,
J. Adams$^{18}$,
J. A. Aguilar$^{12}$,
M. Ahlers$^{22}$,
M. Ahrens$^{50}$,
C. Alispach$^{28}$,
A. A. Alves Jr.$^{31}$,
N. M. Amin$^{42}$,
R. An$^{14}$
% ...
\\

\noindent
% ...
$^{12}$ Universit{\'e} Libre de Bruxelles, Science Faculty CP230, B-1050 Brussels, Belgium \\
% ...
$^{14}$ Department of Physics and Laboratory for Particle Physics and Cosmology, Harvard University, Cambridge, MA 02138, USA \\
% ...
$^{17}$ Department of Physics, Loyola University Chicago, Chicago, IL 60660, USA \\
$^{18}$ Dept. of Physics and Astronomy, University of Canterbury, Private Bag 4800, Christchurch, New Zealand \\
% ...
$^{22}$ Niels Bohr Institute, University of Copenhagen, DK-2100 Copenhagen, Denmark \\
% ...
$^{28}$ D{\'e}partement de physique nucl{\'e}aire et corpusculaire, Universit{\'e} de Gen{\`e}ve, CH-1211 Gen{\`e}ve, Switzerland \\
% ...
$^{31}$ Karlsruhe Institute of Technology, Institute for Astroparticle Physics, D-76021 Karlsruhe, Germany  \\
% ...
$^{42}$ Bartol Research Institute and Dept. of Physics and Astronomy, University of Delaware, Newark, DE 19716, USA \\
% ...
$^{50}$ Oskar Klein Centre and Dept. of Physics, Stockholm University, SE-10691 Stockholm, Sweden \\
% ...
$^{59}$ DESY, D-15738 Zeuthen, Germany \\
% ...

\subsection*{Acknowledgements}

\noindent
USA {\textendash} U.S. National Science Foundation-Office of Polar Programs,
U.S. National Science Foundation-Physics Division,
U.S. National Science Foundation-EPSCoR,
Wisconsin Alumni Research Foundation,
Center for High Throughput Computing (CHTC) at the University of Wisconsin{\textendash}Madison,
Open Science Grid (OSG),
Extreme Science and Engineering Discovery Environment (XSEDE),
Frontera computing project at the Texas Advanced Computing Center,
U.S. Department of Energy-National Energy Research Scientific Computing Center,
Particle astrophysics research computing center at the University of Maryland,
Institute for Cyber-Enabled Research at Michigan State University,
and Astroparticle physics computational facility at Marquette University;
Belgium {\textendash} Funds for Scientific Research (FRS-FNRS and FWO),
FWO Odysseus and Big Science programmes,
and Belgian Federal Science Policy Office (Belspo);
Germany {\textendash} Bundesministerium f{\"u}r Bildung und Forschung (BMBF),
Deutsche Forschungsgemeinschaft (DFG),
Helmholtz Alliance for Astroparticle Physics (HAP),
Initiative and Networking Fund of the Helmholtz Association,
Deutsches Elektronen Synchrotron (DESY),
and High Performance Computing cluster of the RWTH Aachen;
Sweden {\textendash} Swedish Research Council,
Swedish Polar Research Secretariat,
Swedish National Infrastructure for Computing (SNIC),
and Knut and Alice Wallenberg Foundation;
Australia {\textendash} Australian Research Council;
Canada {\textendash} Natural Sciences and Engineering Research Council of Canada,
Calcul Qu{\'e}bec, Compute Ontario, Canada Foundation for Innovation, WestGrid, and Compute Canada;
Denmark {\textendash} Villum Fonden and Carlsberg Foundation;
New Zealand {\textendash} Marsden Fund;
Japan {\textendash} Japan Society for Promotion of Science (JSPS)
and Institute for Global Prominent Research (IGPR) of Chiba University;
Korea {\textendash} National Research Foundation of Korea (NRF);
Switzerland {\textendash} Swiss National Science Foundation (SNSF);
United Kingdom {\textendash} Department of Physics, University of Oxford.

\end{document}